\documentclass{aa}  

\usepackage{graphicx}
\usepackage{amsmath}
\usepackage{amssymb}
\usepackage{txfonts}
\usepackage{threeparttable}
\usepackage[final]{microtype}
\usepackage[dvips]{color}
\usepackage[normalem]{ulem}
\usepackage{fancybox}

\begin{document} 

\title{A non-pulsating neutron star in the supernova remnant
HESS~J1731$-$347 / G353.6$-$0.7 with a carbon atmosphere}

   \author{D.~Klochkov
          \inst{1} \and
          G.~P\"uhlhofer\inst{1}\and
          V.~Suleimanov
          \inst{1,2}\and
         S.~Simon\inst{1}\and
         K.~Werner\inst{1}\and
          A.~Santangelo\inst{1}
          }

   \institute{Institut f\"ur Astronomie und Astrophysik, Universit\"at
     T\"ubingen (IAAT), Sand 1, 72076 T\"ubingen, Germany
     \and
     Kazan (Volga region) Federal University,
     Kremlevskaya 18, 420008 Kazan, Russia 
   }

   \date{Received ***, 2012; accepted ***, 2012}

 
  \abstract
   {The CCO (central compact object) candidate in the center of the supernova remnant shell
     HESS J1731$-$347 / G353.6$-$0.7 shows no pulsations and exhibits a
     blackbody-like X-ray spectrum. If the absence of pulsations is
     interpreted as evidence for the emitting surface area being the
     entire neutron star surface, the 
     assumption of the measured flux being due to a 
     blackbody emission translates into a source distance
     that is inconsistent 
     with current estimates of the remnant's distance.
    }
   {With the best available observational data, we extended
     the pulse period search down to a sub-millisecond time scale and
     used a carbon atmosphere model to describe the X-ray
     spectrum of the CCO and to estimate geometrical parameters of the
     neutron star.
    }
   {To search for pulsations we used data of an observation
     of the source with \emph{XMM-Newton} performed in timing
     mode. For the spectral analysis, we used earlier \emph{XMM-Newton}
     observations performed in imaging mode, which permits a more accurate
     treatment of the background. The carbon atmosphere models used to
     fit the CCO spectrum are computed assuming hydrostatic and
     radiative equilibria and take into account
     pressure ionization and the presence of spectral lines.
    }
   {Our timing analysis did not reveal any pulsations with a pulsed
     fraction above $\sim$8\% down to 0.2\,ms. This finding further
     supports the hypothesis that the emitting surface area is the
     entire neutron star surface. The carbon atmosphere model provides
     a good fit to the CCO spectrum and leads to a normalization
     consistent with the available distance estimates of the
     remnant. The derived constraints on the mass and radius
     of the source are consistent with 
     reasonable values of the neutron star mass and radius. After the CCO in
     Cas~A, the CCO in HESS J1731$-$347 / G353.6$-$0.7 is the second
     object of this class for which a carbon atmosphere model
     provides a consistent description of X-ray emission.
   }
   {}

   \titlerunning{A non-pulsating neutron star in HESS J1731-347 / G353.6-0.7}
   \maketitle

%
\section{Introduction}

The observed population of isolated neutron stars in the Galaxy is
dominated by radio pulsars ($\sim$2000). The class of radio-quiet 
X-ray emitting isolated neutron stars is much less numerous, only
several tens. Members of this class appear in four different flavors:
\emph{anomalous X-ray pulsars} (AXP), \emph{soft gamma-ray repeaters}
(SGR), \emph{central compact objects} (CCO), and \emph{X-ray dim
  isolated neutron stars} (XDINS).
The sources from the former
three classes are in many cases associated with supernova
remnants (SNR) and, thus, are believed to be relatively young
($\sim$$10^3 - 10^4$ yr). Central compact objects are defined through 
their locations near the geometrical center of young SNRs, their steady flux, 
predominantly thermal X-ray emission ($kT\sim 0.2-0.5$\,keV),
and the lack of associated pulsar 
wind nebula (PWN) emission 
\citep[e.g.,][for recent reviews]{Pavlov:etal:2004,Halpern:Gotthelf:10}.
Currently, less than ten such objects are known.
Pulse periods have been measured for three of them and range from 0.1
to 0.4~s. The spin-down rates or the
corresponding upper limits for these three pulsars 
lead to relatively small estimated
magnetic field strengths: $\lesssim$10$^{10}\div 10^{11}$\,G.
These estimates led \citet{Halpern:Gotthelf:10} to name these 
objects anti-magnetars. The 
characteristic age of these pulsars
derived from the $P/\dot P$ ratio
is much larger than the associated 
SNR age, which means that the pulsars were born with spin periods 
close to their present ones, assuming that the magnetic field stayed
constant \citep[see, however,][]{Ho:11,Vigano:Pons:12}. 

The recently discovered CCO candidate \object{XMMUJ173203.3$-$344518}
\citep{Acero:etal:09, Tian:etal:10} is a point-like 
source located close to the 
geometrical center of the shell-type supernova remnant 
\object{HESS~J1731$-$347} or \object{G~353.6$-$0.7}. 
The SNR itself is quite unusual as it was first discovered as a strong 
TeV emitter with HESS \citep{Aharonian:etal:08}, and only subsequently 
was the corresponding weak radio SNR shell identified
\citep{Tian:etal:08}. Thus, the remnant belongs to a relatively rare class of 
TeV-emitting SNRs with only six other TeV-emitting SNRs
known so far \citep[e.g.,][]{Aharonian:2013}. The source
XMMUJ173203.3$-$344518
was discovered during 20\,ksec follow-up observations 
of the SNR with \textsl{XMM-Newton} in 2007 
and was in the FOV during subsequent \textsl{Suzaku}
\citep{Bamba:etal:12} and \textsl{Chandra} observations of the remnant.
The X-ray spectrum of the source can be well
described by an absorbed blackbody model with $kT\simeq 0.5$\,keV
and an absorption column density
of $n_{\rm H}\simeq 1.5\times 10^{22}$\,cm$^{-2}$ 
\citep{Acero:etal:09,Halpern:Gotthelf:10:b,Bamba:etal:12}. 
A possible deviation from a pure blackbody model has been discussed by 
\citet{Tian:etal:10} and \citet{Halpern:Gotthelf:10:b}
and was modeled through an additional higher temperature blackbody or
hard power-law component.  
However, contamination of the point-source spectrum by the surrounding
SNR could be responsible for one or several additional components as
demonstrated in \citet{Bamba:etal:12}, for example.  

So far, no long-term X-ray variability has been detected for
XMMUJ173203.3$-$344518
\citep{Halpern:Gotthelf:10:b}. \citet{Tian:etal:10} 
claimed that the additional hard spectral component was different
in the 2007 \emph{Suzaku} and \emph{XMM-Newton} observations of
the source, but given the uncertainties of the spectral extraction for
this hard component, no strong conclusions can be drawn.

Measurements of X-ray pulsations 
are crucial for the characterization of a CCO. 
So far, observations have not revealed
any hint of pulsations in XMMUJ173203.3$-$344518 down to 10\,ms with
an upper limit for the amplitude of a sinusoidal signal of $\sim$10\%
\citep{Halpern:Gotthelf:10c}. 
The absence of pulsations or a very low pulsed fraction indicate that
the observed thermal radiation of the CCO is emitted by nearly the
entire neutron star surface. 
There is also a chance that the spin axis of
the pulsar points under a small angle towards Earth 
or that the emitting regions are located at
the rotation  poles of the star, which would suppress pulsations.
For this paper, we adopt the hypothesis of a near-complete
neutron star surface emission region.

If thus the entire neutron star surface is the source of
emission, an interpretation of the spectrum as pure blackbody
radiation would place the star at a distance of 30\,kpc 
(for a stellar radius of 10\,km). Such a large distance would imply an
exceptionally high TeV luminosity of the remnant shell compared to other
known TeV-emitting SNRs. 
Together with the SNR's lower distance limit of $\sim$3.2\,kpc
and using the same luminosity argument,
\citet{HESS:2011} argued for a likely localization of
\object{HESS J1731$-$347} either in the Scutum-Crux arm or in the
Norma-Cygnus arm, with the corresponding spiral arm distances of
$\sim$3\,kpc or $\sim$4.5\,kpc following the model by
\citet{Hou:etal:09} for a Galactic center distance of
8.5\,kpc. 

A similar discrepancy between the distance estimates to the SNR shell
and to the associated neutron star 
has been reported for another object, the
non-pulsating CCO in the Cas~A supernova remnant  
\citep[see, e.g.,][and references therein]{Pavlov:Luna:09}, for which the
distance has been measured very reliably
\citep[e.g.,][]{Reed:etal:95}. For Cas~A, a plausible solution of the
discrepancy has been presented by \citet{Ho:Heinke:09}, who fitted the observed
spectra of the CCO using a neutron star carbon atmosphere
model. Their analysis brought the emitting area in agreement with the
total surface area of a canonical neutron star. 

In the present work, we also use a carbon
atmosphere spectral model to reconcile 
the distance estimates of the SNR HESS J1731$-$347 with the emitting
area of the CCO in its center which we have assumed to be the entire
surface of the neutron star. Before doing so,
we used the recent timing
observations of the source with \emph{XMM-Newton}
to confirm the absence of pulsations down to $\sim$0.2\,ms and the
absence of any long-term flux variations, as well. 

The paper is organized as follows. The observation history 
is provided in Sect.\,\ref{sec:var}, where the long-term
stability of the source's flux is discussed. The timing analysis,
including the search for pulsations, is described in
Sect.\,\ref{sec:timing}. The spectral analysis is described in
Sect.\,\ref{sec:spe}. The details of the carbon atmosphere model and
the procedure we used to fit it to the spectrum are described in
Sect.\,\ref{sec:carbon} where we 
also discuss the constraints on the mass and radius of the neutron star. 
The results are summarized in Sect.\,\ref{sec:summary}

\section{Long-term stability of XMMUJ173203.3$-$344518\label{sec:var}}

The CCO in HESS J1731$-$347 has been observed several times with the
\emph{XMM-Newton, Chandra, Suzaku,} and \emph{Swift} orbital
observatories. Table~\ref{tab:obs} provides a short summary of the
observations. Most of them were aimed at the diffuse emission of
the supernova remnant shell and, therefore, were performed in imaging
modes with limited time resolution. The last two observations in
the table were, however, targeted at the CCO and performed in timing
modes providing sufficient time resolution to search for
pulsations down to a millisecond time scale. 
There were several short \emph{Swift} observations of the
source with exposures below 1\,ksec each, which are not listed in
the table.

\begin{table}
  \centering
  \caption{Summary of observations of 
    XMMUJ173203.3$-$344518}
  \label{tab:obs}
  \begin{tabular}{l l l l}
    \hline\hline
    Date & Satellite & exposure & time res.\\
         &           &  [ksec]  &          \\
    \hline
                &                   &    &     \\
    2007 Feb 23 & \emph{Suzaku}     & 41 & 8\,s\\
    2007 Mar 21 & \emph{XMM-Newton} & 25 & 70\,ms (PN)\\
    2008 Apr 28 & \emph{Chandra}    & 30 & 3.2\,s\\
    2009 Feb 4 & \emph{Swift}      & 1.4& 2.5\,s\\
    2009 Mar 9 & \emph{Swift}      & 1.4& 2.5\,s\\
    2010 May 18 & \emph{Chandra}    & 40 & 2.85\,ms\\
    2012 Mar 2 & \emph{XMM-Newton} & 24 & 0.03\,ms (PN)\\
     \hline
  \end{tabular}
\end{table}

The source XMMUJ173203.3$-$344518 is unresolved down to Chandra
resolution. \citet{Halpern:Gotthelf:10:b} reported an X-ray nebulosity
around the point source out to $\sim$50$^{\prime\prime}$, the characteristics of
which are compatible with a scattering halo. From these observations
and the others as well there is no evidence for a pulsar wind nebula 
around the source, in agreement with the CCO hypothesis for
XMMUJ173203.3$-$344518. 

\begin{figure}
\centering
\resizebox{\hsize}{!}{\includegraphics{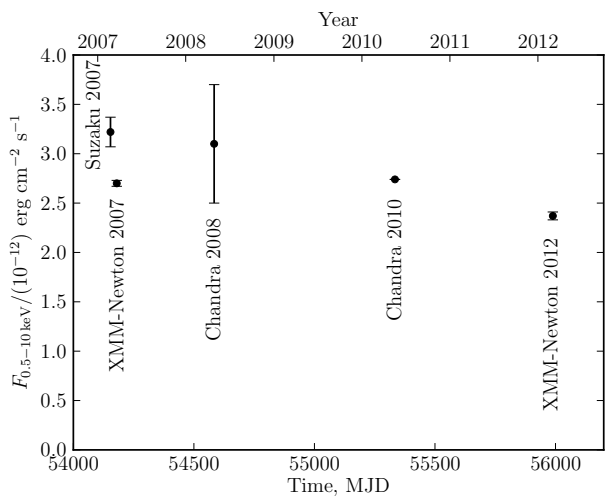}}
\caption{The absorbed flux measurements of the CCO in HESS J1731$-$347 in the
  0.5--10\,keV energy range 
  performed with different instruments. For the data point of
  \emph{Chandra} 2010 taken from \citet{Halpern:Gotthelf:10c}, no error
  bar is available.}
\label{fig:lclong}
\end{figure}

To check for any long-term variations of the source,
we compiled the integrated point source flux measurements derived
from the \emph{XMM-Newton, Chandra}, and \emph{Suzaku} observations in
Fig.\,\ref{fig:lclong}. Fluxes from the 2007 and 2008 data were
derived from a single blackbody
fit. \citet{Halpern:Gotthelf:10:b} already noticed
that the source fluxes from these data (also including the 2009
\emph{Swift} measurements) are consistent within 20\%, and excluded
magnetar-like activity. To the light curve we added the results from
the two timing-mode measurements, the \emph{Chandra} 2010 flux point
derived by \citet{Halpern:Gotthelf:10c} with a two-blackbody
fit as favored by these authors, and the \emph{XMM-Newton} 2012 flux point
as derived from a single blackbody fit.
The new flux measurement,
$F_{\rm 0.5-10\,keV}=2.37(4)\times 10^{-12}$~erg~cm$^{-2}$~s$^{-1}$
(absorbed), indicates a flux that is a few sigma lower than that measured with 
\emph{XMM-Newton} in 2007 and with the other instruments. 

It cannot, however, be excluded
that the systematic differences in the instrument response in
different observing modes (timing 
and imaging) and the systematic uncertainties in the absolute flux 
calibration (for \emph{XMM-Newton} it is believed to be around 
10\%\footnote{\texttt{http://xmm2.esac.esa.int/docs/documents/CAL-TN-0052.ps.gz}})
are responsible for the observed flux differences. We conclude,
therefore, that these differences are not sufficient to claim
long-term variability of the source.
Our results are, therefore, consistent with the CCO hypothesis
for XMMUJ173203.3$-$344518.

\section{Timing analysis\label{sec:timing}}

The observations of the CCO in HESS J1731$-$347 with \emph{Chandra}
in 2010 (see Table~\ref{tab:obs}), performed in timing mode, were used by 
\citet{Halpern:Gotthelf:10c} to search for pulsations down to a period
of 10\,ms. As mentioned in the introduction, the search did not reveal any
pulsations with pulsed fraction larger than $\sim$10\%, assuming a
sinusoidal pulse shape. 

In this work, we analyze the more recent
timing observations of the CCO performed with the \emph{XMM-Newton} 
observatory \citep{Jansen:etal:01} in March
2012 with an exposure time of $\sim$23\,ksec. 
The high time resolution of the EPIC-PN camera in timing mode 
\citep[0.03\,ms,][]{Strueder:etal:01} allows the search 
for coherent pulsations to be extended down to sub-millisecond
periods, that is, down to and beyond the theoretical lower limit of
the neutron star spin period. 

Data processing was performed with the Science Analysis
System (SAS) version 12.0.0. No high-background periods had to be
excluded from the analysis. 
We then used a Rayleigh periodogram, also known as $Z_1^2$-statistics  
\citep[e.g.,][and references therein]{Protheroe:etal:87}, 
to calculate the power spectrum down to the period of 0.2\,ms 
from the event list. 
The maximum value $Z_1^2=36.4$ is below the 
99\% confidence level threshold $Z_{1,99\%}^2=46.3$ and so, no
pulsations are detected at this confidence level.
The threshold is obtained by calculating the $Z_1^2$-power for which the
probability to exceed it by noise is 0.01, and taking into account the
number of trials (in our case the total number of independent
frequencies in the power spectrum, $\sim$1.1$\times
10^{8}$).  
To convert the maximum value of $Z_1^2$ to an upper limit, 
we used the approach described by \citet{Brazier:94} and calculated
theoretical $Z_1^2$-distributions for the case of the presence of a signal.
The resulting upper limit at 99\% confidence level 
for the pulsed fraction of a sinusoidal
pulsation is 8.3\%.

Recently, it has been shown that pulsations of the CCO
\object{RX\,J0822$-$4300} in the SNR Puppis~A exhibit a remarkable
phase shift at the energies slightly above 1\,keV
\citep{Gotthelf:Halpern:09}. This shift would result in a
smearing of pulsations extracted from the data containing photons
with energies above and below the energy of the phase shift. 
The smearing would reduce the sensitivity of our pulsation search
technique. To address this point, we performed separate searches for
pulsations at the photon energies $E$ above and below 1.5\,keV using the
procedures described above. In both cases, no pulsation was found.
The corresponding 99\%\,c.l. upper limits on the amplitude of a
sinusoidal pulsation are 20.0\% for
$E<1.5$\,keV and 9.4\% for $E>1.5$\,keV.

To conclude on the timing analysis, pulsations of the neutron star
from typical CCO periods of a few seconds down to the fastest possible
frequencies can be excluded to below 8.3\% pulsed fraction
assuming that no strong energy-dependent shift of the pulse phase is
exhibited by the source. 

\section{Spectral analysis\label{sec:spe}}

Our spectral study requires a precise measurement of the spectral
continuum of the source with a proper treatment/removal of any
possible background contribution.
Thus, we base our analysis on the 25\,ksec \emph{XMM-Newton}
data from the observations in 2007. We decided not to include the
\emph{Suzaku} data because the limited spatial resolution of the XIS
cameras does not permit an accurate subtraction of the local diffuse
emission generated by the remnant. The 2008 \emph{Chandra} observations are
not used because of the heavy pile-up effect that strongly limits the
accuracy of the spectral modeling. The short \emph{Swift} pointings
provide a negligible statistical contribution to the spectra and are
therefore also omitted. Finally, we do not use the latest \emph{Chandra} and
\emph{XMM-Newton} timing observations because of difficulties with
finding a proper local source-free region for subtraction of the
diffuse emission of the remnant in the two-dimensional slices of the
sky available in timing mode observations. 

The data processing for all three \emph{XMM-Newton} instruments (EPIC-PN and two
EPIC-MOS, \citealt{Turner:etal:01}) was performed with SAS 
version 8.0.0. To extract the source spectrum, we used a
circular extraction region with a radius of $50^{\prime\prime}$
centered at the CCO and encompassing more than 90\% of the PSF. The
background spectrum is extracted from an annulus region around the
source with inner and outer radii of $70^{\prime\prime}$
and $110^{\prime\prime}$ respectively. 
For the spectral extraction, we used the energy
range from 0.5 to 9\,keV.
The spectra from all three cameras are found to be consistent
and were, therefore, fitted simultaneously in the
subsequent analysis. 

The extracted spectrum of the CCO can be well described with an
absorbed blackbody model with best-fit temperature
$kT=0.49(1)$~keV and absorption column density 
$n_{\rm H}=1.50(4)\times 10^{22}$\,cm$^{-2}$. 
The fit yields $\chi^2_{\rm red}/$d.o.f.=1.08/409.
The values are consistent with blackbody values derived from
the same data set by \citet{Halpern:Gotthelf:10:b}, from the
\emph{Suzaku} 2007 data set by \citet{Halpern:Gotthelf:10:b} and 
\citet{Bamba:etal:12}, and from a joint
\emph{XMM-Newton/Suzaku} fit by  
\citet{Tian:etal:10}.

Since the goal of the present work is to fit the spectrum
with the physical neutron star atmosphere model described in the
following section, we refrained from searching for more complex
parametric models which may slightly improve the fit quality.

\section{A carbon atmosphere model for XMMUJ173203.3$-$344518
\label{sec:carbon}}

Generally, an isolated neutron star is expected to be covered by a
thin (centimeter-scale) atmosphere, which, depending on its chemical
composition, substantially modifies the spectrum of the outcoming
thermal radiation. So far, no reliable observational
constraints on the atmospheric composition of isolated neutron stars
have been obtained. Recently, \citet{Ho:Heinke:09} successfully applied
a carbon atmosphere model to the data of the CCO in the Cas~A
supernova remnant. The analysis allowed them to reconcile the emitting 
surface area obtained from the spectral fit with the canonical radius
of the neutron star. Following \citet{Ho:Heinke:09}, we applied
a carbon atmosphere model to our data of the CCO in HESS J1731$-$347.
The lack of pulsations (see Sect.\,\ref{sec:timing}) suggests that the
emitting area of the source is close to the entire surface of the neutron star,
similar to the CCO in Cas~A. We, therefore, maintained
this assumption in our spectral fits. 

\subsection{The carbon atmosphere model}

In this work, we consider relatively cool model atmospheres with
effective temperatures less than 5\,MK. 
Therefore, we believe that Compton scattering is not important
and include only coherent electron scattering. 
We also do not take into account the influence of magnetic field. 
The surface field strength $B$ of the star thus needs to be low 
enough not to affect the model structures and emergent spectra.
It is well known that H-like atoms completely change their structure
in an external magnetic field 
when the energy of an electron on the lowest Landau level
$\hbar\omega_c/2$ ($\omega_c$ is the cyclotron frequency) 
becomes comparable to the ionization energy for the ground state
or, equivalently, when $B$ becomes comparable to 
$B_{\rm cr} = m_{\rm e}^2 e^3cZ^2/\hbar^3 \approx 2.4\cdot
10^9\,Z^2$\,G \citep[see, e.g.,][]{Harding:Lai:06}. For carbon, 
$B_{\rm cr}\approx 8.5\cdot 10^{10}$~G. We believe that the magnetic
field below (0.001 -- 0.01)$\times$$B_{\rm cr} \approx 10^8 - 10^9$~G 
will not noticeably change the ionization energies and number densities 
of the carbon ions. Thus, we believe
that our models are applicable for $B\lesssim 10^8 - 10^9$\,G.
Since no pulsations are detected in XMMUJ173203.3$-$344518, no
estimates of the surface $B$-field strength are
available. The results of our modeling are thus valid under the
assumption that the surface magnetic field strength of the neutron star
is below the specified limit.

The basic assumptions of the computation of stellar atmosphere models
with coherent electron scattering are described in \citet{Mihalas:78}.
Here, we emphasize the differences caused by carbon domination in the
chemical composition and high density of the neutron star atmospheres.
The models are computed assuming hydrostatic and radiative
equilibria in plane-parallel approximation. 
The main input parameters are the effective
temperature $T_{\rm eff}$, the chemical composition 
(in this work, we consider a pure carbon atmosphere), and
the surface gravity
\begin{equation}
  g=\frac{GM}{R^2}(1+z),
  \label{eq:gdef}
\end{equation}
where $M$ and $R$ are the mass and radius of the neutron star
respectively and $G$ is the gravitational constant. The gravitational
redshift $z$ on the star 
surface is related to the neutron star parameters as
\begin{equation}
  1+z=(1-2GM/c^2R)^{-1/2} .
  \label{eq:zdef}
\end{equation}

\begin{figure}
\centering
\resizebox{0.9\hsize}{!}{\includegraphics{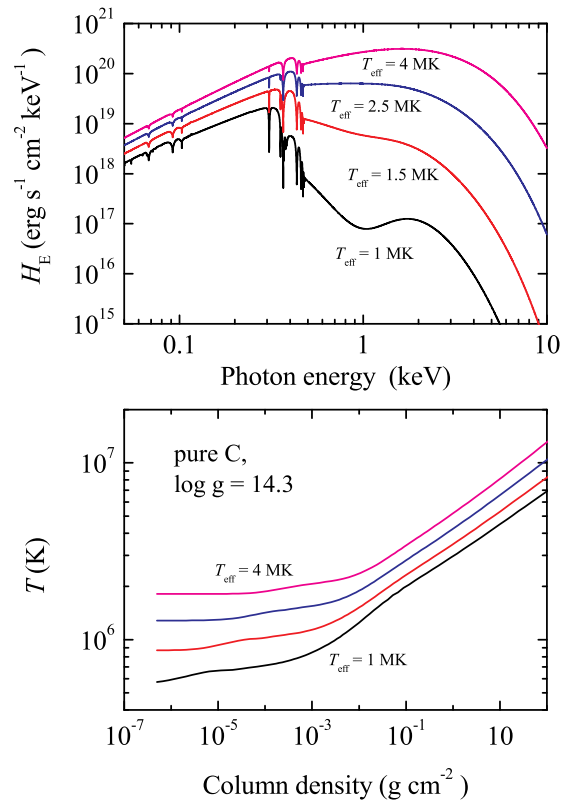}}
\caption{Emergent spectra (top) and temperature
  structures (bottom) of pure carbon atmospheres with a fixed 
  $\log g$ = 14.3 and a set of effective temperatures.}
\label{fig:af1}
\end{figure}

In our calculations we
assume local thermodynamic equilibrium (LTE); therefore,  
the number densities of all ionization and excitation states of carbon
are calculated using the Boltzmann and Saha equations. 
We accounted for the pressure ionization effects on carbon populations
using the occupation probability formalism \citep{Hummer:Mihalas:88},
as  described by \citet{Hubeny:etal:94}. In addition to electron
scattering, we took into account the free-free opacity, as well as the
bound-free transitions for all carbon ions using opacities from
\citet{Verner:etal:96} and \citet{Verner:Yakovlev:95}
\citep[see][]{Ibragimov:etal:03}. Line
blanketing is taken into account using carbon spectral lines from the
CHIANTI atomic database, Version 3.0\citep{Dere:etal:97}. 

For our computations we used the numerical
code ATLAS \citep{Kurucz:70,Kurucz:93} modified to deal with high
temperatures
\citep{Ibragimov:etal:03,Suleimanov:Werner:07,Rauch:etal:08}.  

Examples of emergent spectra and temperature structures are
shown in Fig.\,\ref{fig:af1}. Absorption edges
due to the ions CV and CVI are seen in the low-temperature spectra. At
higher temperatures, the number density of CV becomes very small, so that
only the CVI edge is seen. It is clearly visible that the edges change the
spectral continuum dramatically, so that it strongly deviates
from the blackbody spectrum with the corresponding effective
temperature or from a spectrum of a hydrogen model atmosphere
\citep[see][]{Ho:Heinke:09}.   
Our method of the carbon atmosphere modeling 
will be discussed in detail in a separate paper.


\subsection{Fitting the model to the spectra}

The atmospheres described above are computed for a series of
$\log g $ values from 13.70 to 14.90 with a step of 0.15 (the
surface gravity $g$ is
expressed in cgs units) and effective temperatures $T_{\rm eff}$ from
1 to 4 MK with a step size of 0.05 MK. The photon fluxes have been
computed over an energy grid
from 0.01 to 20.00 keV with a step size of 0.02 keV sufficiently
oversampling the spectral resolution of the instrument. The models are
converted into a two-dimensional FITS table that can be imported as an
\texttt{atable} model component in the XSPEC or 
ISIS\footnote{\texttt{http://space.mit.edu/cxc/isis/}}
spectral fitting packages. Both XSPEC and ISIS allow a linear
interpolation between the values in the table, in our case between the
effective temperatures and surface gravities, which is found to
work very smoothly in our spectral fitting. Additionally, XSPEC and
ISIS allow the gravitational redshift $z$ and a
normalization factor $K$ to be included to fitting parameters. 

\begin{figure}
\centering
\resizebox{\hsize}{!}{\includegraphics[angle=-90]{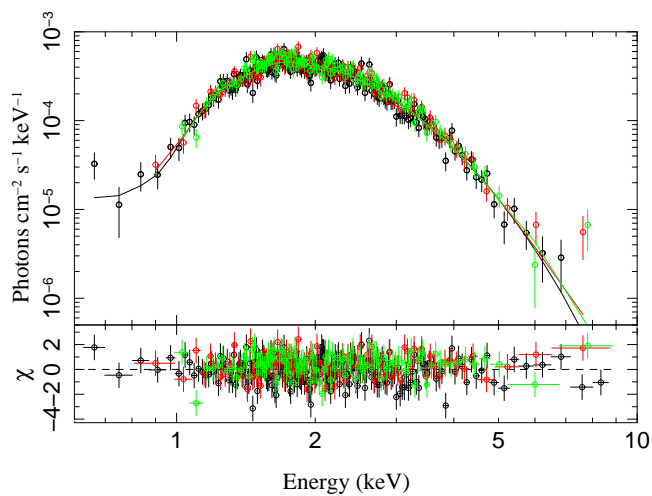}}
\caption{The fit to the spectra of the CCO in HESS J1731$-$347
  obtained with the PN, MOS1, and MOS2 cameras of \emph{XMM-Newton} with
  the carbon atmosphere model assuming a distance of 3.2 kpc.}
\label{fig:spefit}
\end{figure}

In our spectral modeling, we assumed that the carbon atmosphere covers
the entire neutron star 
located at a distance $D$. Taking into account
Eqs.\,(\ref{eq:gdef}) and (\ref{eq:zdef}), 
this implies the coupling between the fit parameters 
\begin{equation}
  K = (r_5/d_{10})^2,
\label{eq:coupling1}
\end{equation}
\begin{equation}
  1 + z = [1-m(0.337r_5)^{-1}]^{-1/2},
\end{equation}
\begin{equation}
  \log g = 16.125+\log [mr_5^{-2}(1+z)],
\label{eq:coupling3}
\end{equation}
where $K$ is the normalization,
$m$ and $r_5$ are the mass and the radius of the neutron star in
units of  solar masses and $10^5$\,cm, respectively, and $d_{10}$ is the
distance to the star in units of 10\,kpc.
The model is multiplied by 
a component accounting for interstellar absorption characterized
by the equivalent column density of hydrogen atoms $n_{\rm H}$.
As described in the introduction, the source
is most probably located either in the Scutum-Crux arm ($\sim$3\,kpc) or 
in the Norma-Cygnus arm ($\sim$4.5\,kpc). The lower limit for the distance
derived by \citet{HESS:2011}
on the basis of the X-ray absorption
pattern and the absorbing column densities based on CO observations is
$\sim$3.2\,kpc. We thus performed our spectral fitting for the two
following fixed
distances, 3.2 and 4.5\,kpc, and assumed the coupling between the
parameters specified in Eqs.(\ref{eq:coupling1})--(\ref{eq:coupling3}).
For both distances, the model provides a
good fit to the data (Fig.~\ref{fig:spefit}). 
The best-fit parameters obtained for the two distances are summarized
in Table~\ref{tab:fitpar}. 

\begin{table}
  \centering
  \renewcommand{\arraystretch}{1.3}
  \caption{Results of the spectral fit of the \emph{XMM-Newton}
    data from the CCO in HESS J1731$-$347 with the carbon atmosphere
    model described in the text for two fixed distances. The $P$-value is
    the null hypothesis probability associated to the respective
    $\chi^2$ minimum.} 
  \label{tab:fitpar}
  \begin{tabular}{l l l}
    \hline\hline
    Parameter & $D=3.2$\,kpc & $D=4.5$\,kpc \\
    \hline
                &            &     \\
    $n_{\rm H}/(10^{22}\,{\rm atoms\,cm}^{-2})$ & $1.95_{-0.07}^{+0.06}$ & $1.95_{-0.07}^{+0.05}$ \\
    $T_{\rm eff}/(10^6\,{\rm K})$ & $2.2_{-0.2}^{+1.2}$& $2.4_{-0.4}^{+0.9}$ \\
    $m$ & $1.5_{-0.6}^{+0.4}$ & $2.2_{-0.9}^{+0.3}$\\
    $r_5$ & $12.6_{-5.3}^{+2.1}$ & $15.6_{-5.3}^{+3.6}$ \\
    $\chi^2_{\rm red}$/d.o.f. & 0.98/408 & 0.98/408\\
    $P$-value & 0.59 & 0.59 \\
     \hline
  \end{tabular}
\end{table}

\subsection{Constraints to the mass and radius of the neutron star}

\begin{figure}
\centering
\resizebox{\hsize}{!}{\includegraphics{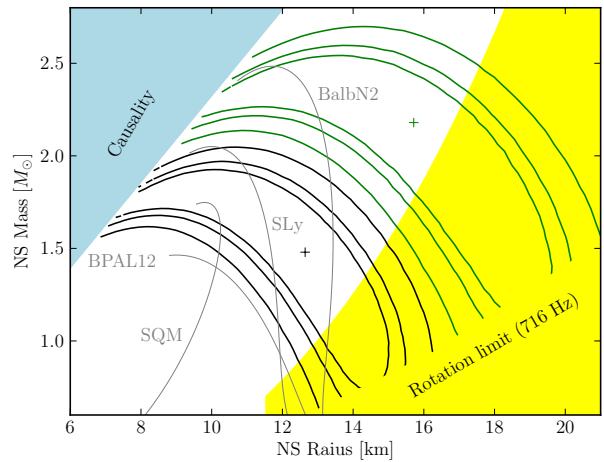}}
\caption{The $\chi^2$ confidence regions in the neutron star mass and
radius plane for the CCO in HESS~J1731$-$347 obtained with the
carbon atmosphere models. The plotted contours correspond to the 50,
68, and 90\% confidence level. The black contours (bottom left)
correspond to the fixed distance of 3.2\,kpc, the green contours (top
right) to 4.5\,kpc. The crosses indicate the corresponding
$\chi^2$-minima. The shaded areas in the top left and bottom right
of the plot indicate the regions excluded by the requirements of
causality and from the fastest rotation-powered pulsar known so far, 
PSR J1748$-$2446ad (716~Hz).} 
\label{fig:RM}
\end{figure}

\begin{figure}
\centering
\resizebox{\hsize}{!}{\includegraphics{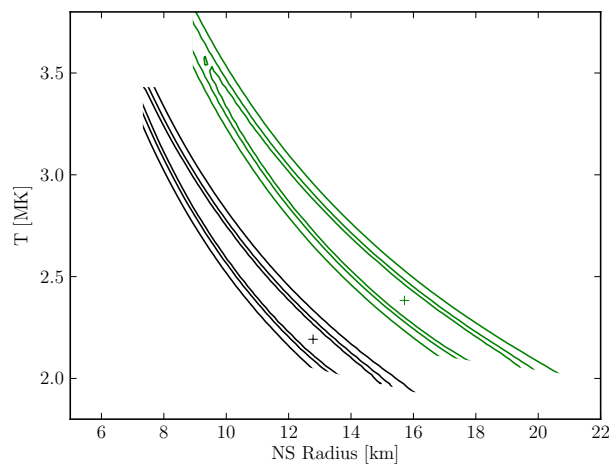}}
\caption{The $\chi^2$ confidence regions for the CCO in
  HESS~J1731$-$347 similar to those in
  Fig.\,\ref{fig:RM} but in the mass -- effective temperature plane.
  As before, the black contours (bottom left)
correspond to the fixed distance of 3.2\,kpc, the green contours (top
right) to 4.5\,kpc.} 
\label{fig:RT}
\end{figure}

Using the carbon atmosphere model, we have explored the $\chi^2$-map
in the neutron star mass--radius plane. The $\chi^2$-contours
at 50, 68, and 90\% confidence level for the two fixed distances are
shown in Fig.~\ref{fig:RM}. 
The shaded area in the bottom right (rotation limit) is excluded for
neutron stars with a similar equation of state as that of the fastest
rotation-powered pulsar known so far, PSR~J1748$-$2446ad (716~Hz).
Both $\chi^2$-minima are almost
equally deep (see Table~\ref{tab:fitpar}). The fit is, thus,
not very sensitive to the distance. 
Both contours favor somewhat larger mass and radius of the 
star compared to the canonical values of 1.4\,$M_\odot$ and 10\,km,
respectively (by a
factor of $\sim$20\% or $\sim$50\% for 3.2 and 4.5\,kpc, respectively).
The 3.2\,kpc contour is, however, compatible with most
nuclear equations of state. We conclude, therefore, that the
spectrum of the CCO in HESS~J1731$-$347 is consistent with that of a
roughly isotropically emitting neutron star 
possessing a carbon atmosphere with an effective temperature of
$\sim$2.3$\times 10^{6}$\,K located at a distance of
$\sim$3--4\,kpc. The source is then very similar to the CCO in the
center of the Cas~A remnant.

If we fix the neutron star mass to the canonical value of
1.4$M_\odot$, the best-fit values of the neutron star radius and the
associated 1$\sigma$ uncertainties are 
$12.7_{-0.9}^{+1.3}$\,km for the distance of 3.2\,kpc and  
$18.4_{-1.1}^{+1.4}$\,km for 4.5\,kpc, respectively. In
the case of 4.5\,kpc, however,
the value lies deep beyond the rotation limit set by the fastest
rotation-powered pulsar PSR J1748-2446ad (716~Hz)
(see Fig.\,\ref{fig:RM}).

Figure\,\ref{fig:RT} shows the $\chi^2$ confidence contours similar to
those in Fig.\,\ref{fig:RM} but in the neutron star radius --
effective temperature plane. The contours indicate strong degeneracy
between the two parameters which is qualitatively
well understood. Indeed, a
higher neutron star radius corresponds to a larger emitting surface
area and, thus, to a higher flux. To reduce the model flux to match the
observed value, one needs to reduce the effective temperature, which
leads to the observed degeneracy. The gravitational redshift
becomes lower with increasing neutron star radius, which compensates for
the softening of the spectrum because of the reduction of the effective temperature.   

We note that the shape of the contours presented in 
Figs.\,\ref{fig:RM} and \ref{fig:RT} might be affected by a
number of effects that are not taken into account in our modeling.
These effects
are (i) a possible influence of the magnetic field; (ii) the
presence of hot spots on the neutron star surface, whose geometry
and location, however, prevent the detection of pulsations; and (iii) the actual
chemical composition of the stellar atmosphere deviating from pure
carbon.

\section{Summary and conclusions\label{sec:summary}}

We performed timing and spectral analyses of the X-ray data from the CCO
candidate XMMUJ173203.3$-$344518 located in the center of the 
shell-type SNR
HESS~J1731$-$347 obtained with the \emph{XMM-Newton} observatory. The
data obtained in timing mode allowed X-ray pulsations of
the source with an amplitude above 8.3\% to be excluded
down to a period of 0.2\,ms,
that is, well beyond the theoretical lower limit of the neutron star
spin period. Based on this finding, we conclude that, unless the spin
axis of the neutron star is nearly co-aligned with the observer's line of
sight or the emitting regions are located roughly at the
star's  rotation poles, 
the observed X-ray emission must be almost uniformly generated by
the entire surface of the neutron star. 

A fit to the X-ray spectrum of XMMUJ173203.3$-$344518 with a pure
blackbody emission model, assuming the canonical neutron star radius
of 10\,km, results in 
an estimated distance to the source of $\sim$30\,kpc, which is
difficult to reconcile with the observed properties of the supernova
remnant HESS~J1731$-$347. The distance to the remnant is estimated to
be $\sim$3.2$-$4.5\,kpc. In this work, we demonstrate that the X-ray
spectrum of the CCO candidate can be described with a pure carbon
atmosphere model under the assumption that the surface
magnetic field of the star does not exceed $\sim$10$^{8}-10^{9}$\,G. 
The spectral fit is consistent with a distance to
the neutron star of $\sim$3.2--4.5\,kpc and a neutron star radius of
$\sim$10--15\,km. We note that a plausible solution of a similar
discrepancy in case of the CCO in Cas~A has recently been
presented by \citet{Ho:Heinke:09} who 
used a carbon atmosphere model to describe the source X-ray
spectrum. We thus conclude that, similarly to the CCO in   
Cas~A, XMMUJ173203.3$-$344518 is most probably a thermally
emitting neutron star with an emitting area uniformly distributed over
the entire neutron star surface covered by a carbon atmosphere.

\begin{acknowledgements}
VS acknowledges the support by the German Research Foundation
(DFG) grant SFB/Transregio 7 "Gravitational Wave Astronomy" and
Russian Foundation for Basic Research (grant
12-02-97006-r-povolzhe-a). We also thank Dr. Craig Heinke for
useful comments and suggestions.
\end{acknowledgements}

\bibliographystyle{aa}
\bibliography{refs}

\end{document}